\documentclass[]{aa}      
\usepackage{graphicx}

\usepackage{txfonts}

\usepackage{multirow}

\usepackage{natbib}
\usepackage{placeins}
\usepackage{bm}
\usepackage{upgreek}

\usepackage[bookmarks=false, colorlinks=true, citecolor=blue, linkcolor=blue]{hyperref}
\usepackage[normalem]{ulem}

\def\kms{\,km\,s$^{-1}$\,}

\begin{document} 

\title{Seven-year periodic variations in the methanol maser line displayed by the massive protostar IRAS\,20216+4104\thanks{The reduced spectra are available at the CDS via anonymous ftp to cdsarc.cds.unistra.fr (130.79.128.5) or via https://cdsarc.cds.unistra.fr/viz-bin/cat/J/A+A/.../...}
}

 \author{M. Szymczak
          \inst{1} \href{https://orcid.org/0000-0002-1482-8189}{\includegraphics[scale=0.5]{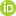}}
          \and
          M. Durjasz
          \inst{1} \href{https://orcid.org/0000-0001-7952-0305}{\includegraphics[scale=0.5]{orcid.png}}
          \and
          S. Goedhart
          \inst{2} \href{https://orcid.org/0000-0003-3636-8731}{\includegraphics[scale=0.5]{orcid.png}}
          \and
          P. Wolak
          \inst{1} \href{https://orcid.org/0000-0002-5413-2573}{\includegraphics[scale=0.5]{orcid.png}}
          \and
          A. Bartkiewicz
          \inst{1} \href{https://orcid.org/0000-0002-6466-117X}{\includegraphics[scale=0.5]{orcid.png}}
          \and
          A. Caratti o Garatti
          \inst{3} \href{https://orcid.org/0000-0001-8876-6614}{\includegraphics[scale=0.5]{orcid.png}}
          \and
          A. Kobak
          \inst{1} \href{https://orcid.org/0000-0002-1206-9887}{\includegraphics[scale=0.5]{orcid.png}}
          \and
          F. Massi
          \inst{4} \href{https://orcid.org/0000-0001-6407-8032}
          {\includegraphics[scale=0.5]{orcid.png}}
          } 
  
   \institute{Institute of Astronomy, Faculty of Physics, Astronomy and Informatics, Nicolaus Copernicus University, Grudziadzka 5, 87-100 Torun, Poland
    \and
    South African Radio Astronomy Observatory, 2 Fir Street, Observatory 7925, South Africa
    \and
    INAF - Osservatorio Astronomico di Capodimonte, salita Moiariello 16, 80131, Napoli, Italy
    \and
    INAF - Osservatorio Astrofisico di Arcetri, Largo E. Fermi 5, I-50125 Firenze, Italy
}

\date{Received 9 October 2023/ Accepted 10 November 2023}

 
  \abstract
  {}
{We report the discovery and analysis of a periodic methanol maser in the massive protostar IRAS\,20216+4104.}
{To obtain the light curve, we used the 6.7\,GHz methanol maser spectra collected between 2000-2003 and 2009-2023 with the Hartebeesthoek and Torun radio telescopes, as well as spectra from the literature reported prior to 1992.}
{The velocity-integrated flux density shows sinusoidal-like variations with a period of 6.9$\pm$0.03\,yr. All but one of the features show periodic changes with a relative amplitude of 2 up to $>$89. A slightly variable feature displays a moderate anti-correlation between the flux density and the other significantly variable features. The maser emission appears to follow the continuum emission of the red-shifted outflow cavity. A maximum emission of 3.4 and 4.6\,$\mu$m precedes the maser peak by 15\% of the period and the (infrared) IR light centroids show time-dependent displacement. The periodic behaviour of the maser and IR emission is likely due to the eclipsing effect from a wobbling inner disk.
}
{}


\titlerunning{Periodic variations of methanol maser in IRAS\,20216+4104}
\authorrunning{M. Szymczak et al.}

\maketitle

\section{Introduction}
Astrophysical masers are known to be very sensitive to changes in the physical conditions in their host clouds, and primarily those caused by radiation and collision processes. For class II methanol masers, diverse variability patterns occur from very complex alteration of the spectra with the appearance and disappearance of features, bursts of individual features and their groups on a timescale of a few days to years to a steady rise and/or decline in feature peaks over several years (e.g., \citealt{goedhart2004}; \citealt{szymczak2018a}). While most methanol maser sources do not show predictable variability patterns, there is a growing group of 6.7\,GHz methanol maser sources with strictly periodic variability. About 30 sources with period ranging from 24 to 1260\,d with a median of 220\,d have been discovered so far (\citealt{araya2010}; \citealt{goedhart2003,goedhart2004, goedhart2009, goedhart2014}; \citealt{fujisawa2014}; \citealt{maswanganye2015,maswanganye2016}; \citealt{olech2019,olech2022}; \citealt{proven-adzri2019}; \citealt{sugiyama2015,sugiyama2017}; \citealt{szymczak2011,szymczak2015,szymczak_2016}; \citealt{tanabe2023}).

The case of a wide span of periods and diversity of light curves from sinusoidal-like to intermittent bursts with a relative amplitude of 0.2 to more than 120 has brought on competing hypotheses on the nature of the maser periodicity phenomenon. \cite{van_der_walt2011} (see also \citealt{van_der_walt2016}) proposed changes in the seed photon flux due to the orbital modulation of the free-free emission of the hot post-shock gas in a colliding-wind binary (CWB) system as a primarily periodicity mechanism. This simple model satisfactorily explains the maser light curves with an asymmetric flare profile, namely, with a fast rise and slow decay in flux density, which is observed in about a half of the periodic 6.7\,GHz masers (\citealt{szymczak2015}). A more sophisticated model of a CWB aptly describes the flare profile of several sources and light curves over a decade (\citealt{vanderheever2019}). 

Methanol maser periodicity can be driven by luminosity variations in a binary system with a massive primary (\citealt{araya2010,parfenov2014}) or the pulsation of a single massive star (\citealt{inayoshi2013}), which affect the dust temperature and, consequently, the pumping rate. Luminosity variations can be due to periodic modulations of accretion rate in binary systems (e.g., \citealt{artymowicz1996,munoz2016,parfenov2014}). In contrast to the CWB model, these mechanisms do not predict a specific flare profile or light curve. \cite{gray2020} performed 3D model simulations for a maser cloud, ultimately concluding that a flare profile driven by a change in pump rate is qualitatively different from that caused by background variation. After applying the observed IR light curves of two sources as the driving function in their model, they rejected the variable background mechanism for the flaring 6.7\,GHz methanol masers. 

A scarcity of long ($>$2\,yr) period masers may be due to the shortness of monitoring observations, which usually lasts 3-4\,yr (\citealt{goedhart2004}; \citealt{szymczak2018a}) because in those programmes, a periodicity that is on timescales of less than 220\,d could be sampled effectively. The recent finding of 1260\,d periodicity in G5.900$-$0.430 has resulted from a 10\,yr monitoring (\citealt{tanabe2023}). Here, we report the 6.7\,GHz maser periodicity in IRAS\,20126+4104 that is longer by a factor of 2.

IRAS\,20216+4104 is one of the most frequently studied high-mass young stellar objects (HMYSOs) with the disc-outflow system (\citealt{cesaroni2007} for review).  The central protostar has a mass of $\sim12M_{\odot}$ (\citealt{cesaroni2014, cesaroni2023}; \citealt{chen2016}) and a luminosity of $\sim10^4L_{\odot}$ (\citealt{johnston2011}; \citealt{cesaroni2023}) at distance of 1.64\,kpc (\citealt{moscadelli2011}).  A resolved accretion disc traced by several molecular lines undergoes Keplerian rotation (\citealt{cesaroni2005}), but its structure has no axial symmetry in the CH$_3$CN(12–11) K=8 line emission, probably due to the effect of nearby stellar companions (\citealt{cesaroni2014}; \citealt{massi2023}). Multi-epoch photometry of the outflow cavities reveals possibly periodic variability of the NIR continuum that may be due to the wobbling of the inner disc caused by a nearby low-mass star companion (\citealt{massi2023}). The water maser emission at 22\,GHz traces a bipolar outflow of velocity up to $\sim$100\kms (\citealt{moscadelli2011}). A sparsely sampled light curve of the 22\,GHz water maser from 1989 to 2007 is reported in  \cite{felli2007}; it does not show any apparent periodicity or trends. 

The 6.7 GHz methanol maser emission does not delineate any disc-like structure; it has appeared only in the northern parts of the disc at 390-530\,au from a central star whose position was derived from the water maser modelling  (\citealt{moscadelli2011}; \citealt{cesaroni2013}). The methanol maser at 6.7\,GHz was observed at two non-overlapping 3-4\,yr intervals (\citealt{goedhart2004}; \citealt{szymczak2018a}). \cite{goedhart2004} reported that the source may be variable, but the signal-to-noise ratio was low. We note, however, a weak variability of $-$6.1\kms\, feature in both studies. Here, we use another data set to get a longer light curve.  

\section{Data}
The HartRAO data used here were taken as part of a monitoring programme published in \cite{goedhart2004}, spanning from January 2000 to February 2003 (MJD 51567$-$52681). The target was observed on average twice a month with a spectral resolution of 0.112\kms and 3$\sigma$ sensitivity of 1.2\,Jy. We obtained data from the Torun radio telescope between June 2009 and September 2023 (MJD 55006$-$60208) at a weekly cadence. The data from the initial period (June 2009 to February 2013) were published in \cite{szymczak2018a}, where the instrument, observations, and data reduction methods were described in detail. In short, a spectral resolution was 0.09\kms, a typical 3$\sigma$ sensitivity of 0.9\,Jy and calibration accuracy of $\sim$10\%

In addition, we searched for observations of the source in the literature and found eight data points spanning from October 1991 until June 2016. The dates, velocity-integrated flux density and instruments are listed in Table\,\ref{tab:data-literature}. Thus, the time range of observations of the target was extended to 31.9 years. Since these spectra were obtained with different sensitivity and resolution, we used the series of integrated flux densities across the entire velocity range
and time interval. We used more homogeneous data obtained with the Torun telescope to analyze the light curve of individual features.

\section{Results}
\begin{figure}
\centering
\includegraphics[width=1\columnwidth]{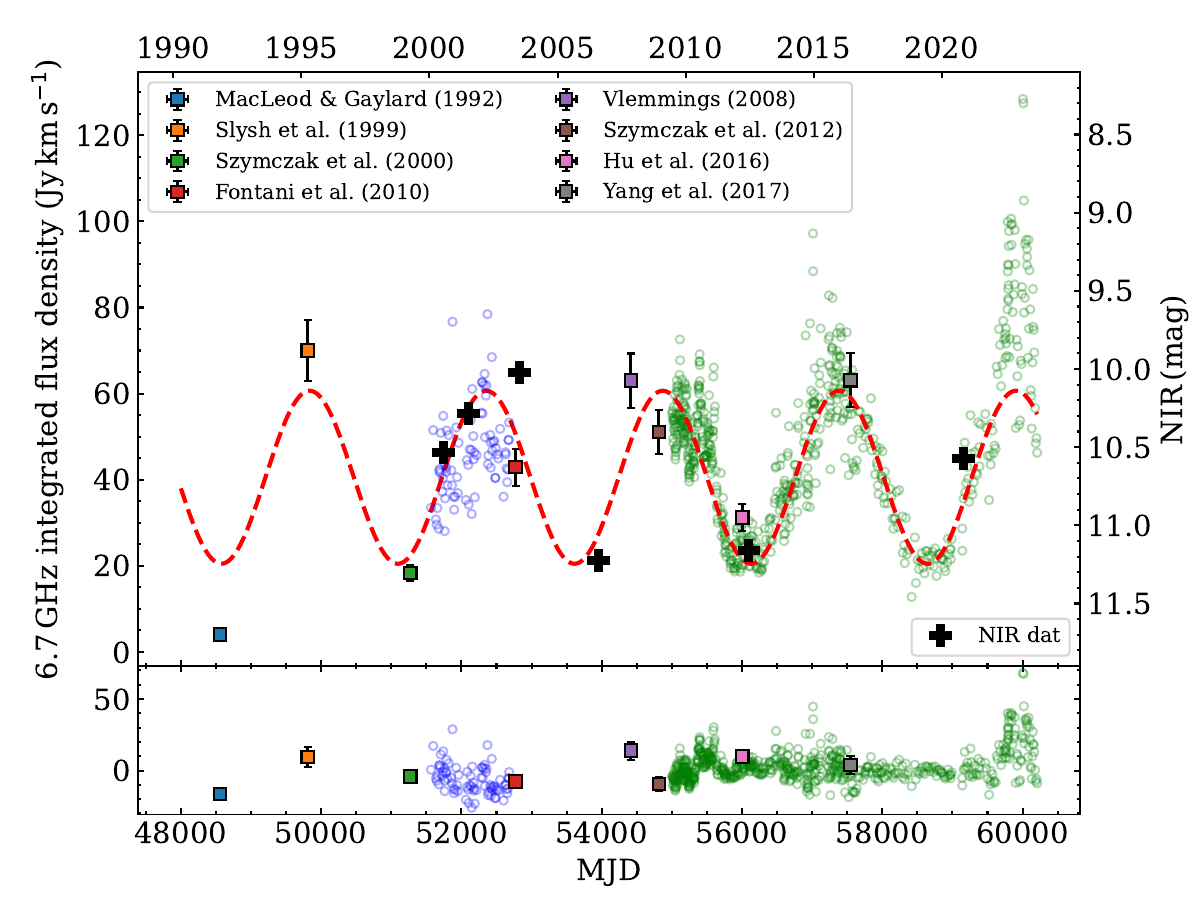}
\caption{Time series of the velocity integrated 6.7\,GHz maser flux density in IRAS\,20216+4104 with the sinusoidal fitted dashed red line (top). The open circles denote the data from the HartRAO (in violet) and Torun (in green) radio telescopes. The squares mark the data from the literature (Table\,\ref{tab:data-literature}) as indicated in the legend. Residuals from subtracting the sinusoidal fitting from the observational data are shown in the bottom panel. The black crosses mark the NIR continuum emission of the south-eastern outflow cavity (\citealt{massi2023}, polygons S1 and S2 in their Fig.\,2).
\label{fig:g78-lc}}
\end{figure}

\begin{figure}
\centering
\includegraphics[width=1\columnwidth]{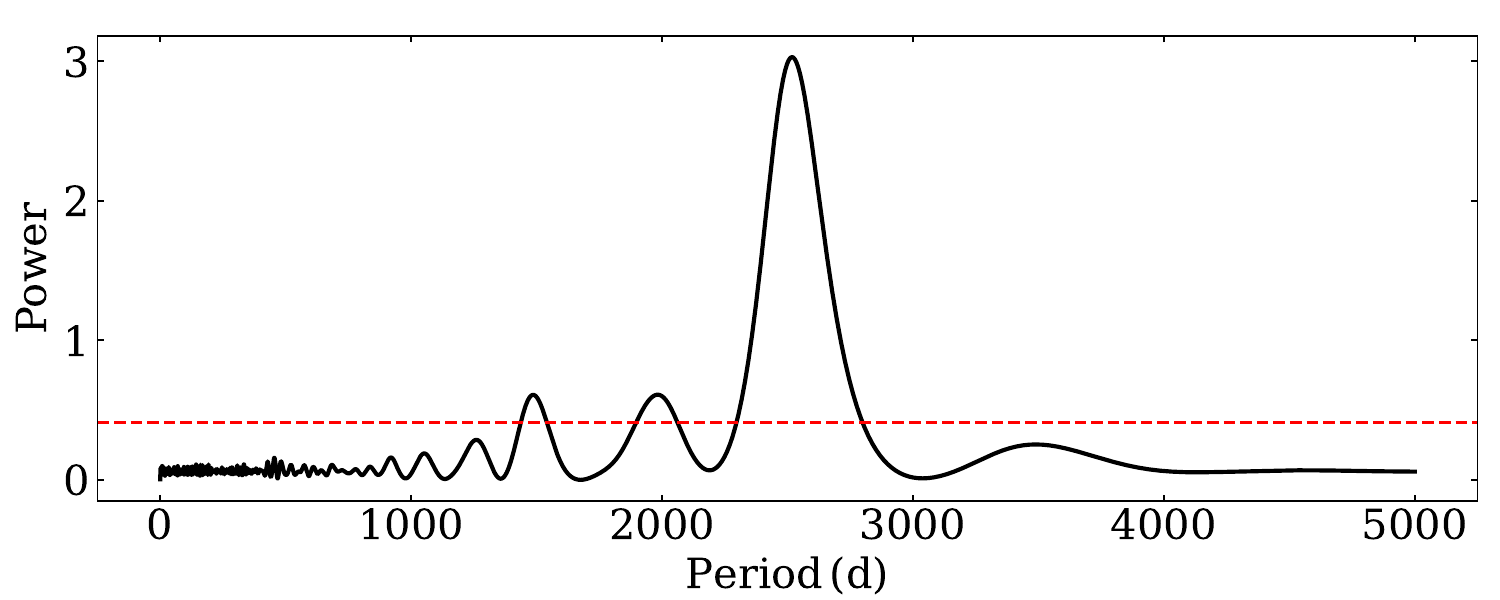}
\caption{Lomb-Scargle periodogram for the time series presented in Fig.\,\ref{fig:g78-lc}. The dashed line represents the 0.1\% false alarm probability level.
\label{fig:g78-LS-perio}}
\end{figure}

Figure\,\ref{fig:g78-lc} shows the maser light curve of IRAS\,20216+4104 over 31.9\,yr and the best fit with a sinusoidal function. Noticeably, the observational data, scattered over time and from different independent instruments, are consistent with the pattern of variability displayed since 2009. The Lomb-Scargle periodogram (Fig.\,\ref{fig:g78-LS-perio}) indicates that the period of variability is 2520$\pm$70\,d, the uncertainty is estimated as the half width at half maximum of the Gaussian function fitted to the top half of the fundamental peak.

Since June 2009 (MJD 55006), the object was observed weekly with four gaps lasting more than two months. The overall picture of variability over 14.2\,yr is shown in Fig.\,\ref{fig:dyn-spec}. 
There is a clear indication (Fig.\,\ref{fig:g78-lc}) that the average flux in the last four maxima increases by about 15\% in each successive cycle, while at the minima, the flux is probably constant. Similar changes of the flare amplitude fitted with a low-degree polynomial were observed, over 3-8\,yr, for some spectral features of periodic (0.4$-$0.6\,yr) masers such as G30.400$-$0.296 and G108.76$-$0.99 (\citealt{olech2019}).

\begin{figure}
\centering
\includegraphics[width=1\columnwidth]{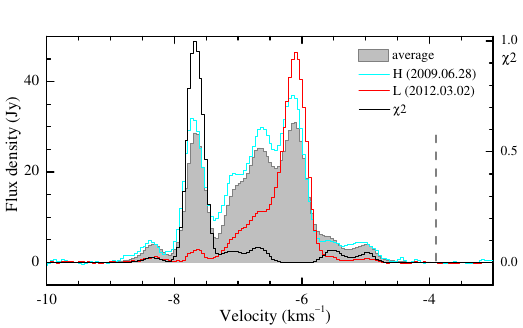}
\caption{Comparison of 6.7\,GHz maser spectra obtained with the Torun telescope. The average spectrum is for 669 observations from MJD 55006 to 60208. The spectra marked by the cyan and red lines are typical for high (H) and low (L) activity states, respectively. The black line represents the normalized $\chi ^{2}$ values as a measure of variability. The dashed vertical line indicates the systemic velocity (\citealt{cesaroni2005}). \label{fig:g78-spec}}
\end{figure}

\begin{figure}
\vspace{-0.5cm}
\includegraphics[width=1\columnwidth]{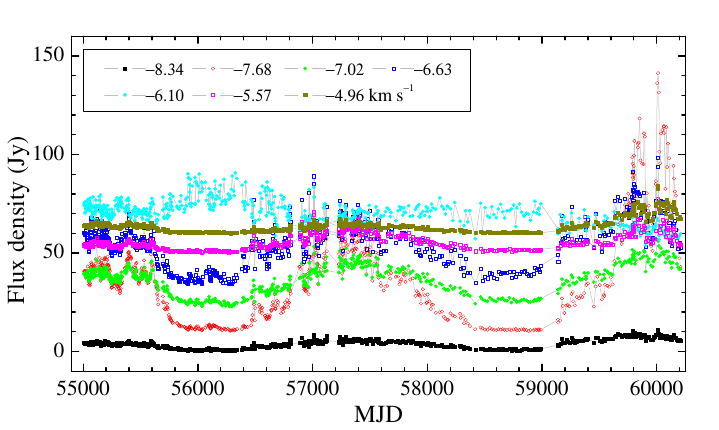}
\caption{Light curves of the main maser features in IRAS\,20216+4104 taken with the Torun telescope. The features are labeled with the peak velocity. The flux density scale of the subsequent curves is shifted by 10\,Jy.
\label{fig:g78-lc-features}}
\end{figure}

\begin{figure}
\includegraphics[width=1\columnwidth]{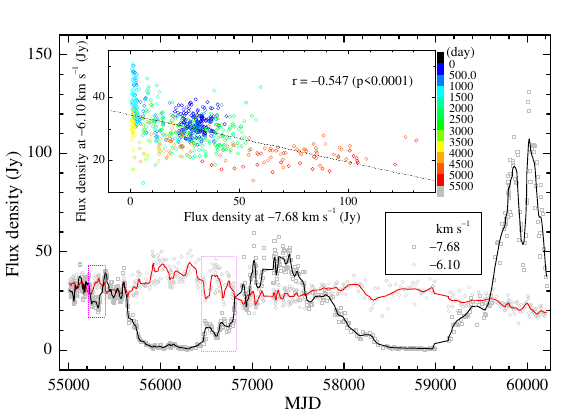}
\caption{Light curves of the $-$7.68 and $-$6.10\kms features. The thick lines show the average curves. The magenta boxes mark the intervals of anti-correlated short-term variability. The inset shows the anti-correlation of these features' flux density, where the symbol's colour denotes the time since MJD 55000.
\label{fig:g78-anti-correlation}}
\end{figure}

The average and typical spectra in the high and low emission states as well as the normalized $\chi^2$ plot are displayed in Fig.\,\ref{fig:g78-spec}. Significant differences exist in the spectra shapes in the low and high states. The light curves of seven features are presented in Fig.\,\ref{fig:g78-lc-features} and their variability properties are listed in Table\,\ref{tab:var-prop}. Except for the $-$6.1\kms feature, all six others show a sinusoidal-like variability with a plateau minimum equally lasted 1.5$\pm$0.2\,yr  in both cycles. In the plateau state, the flux density dropped below our detection limit ($\sim$0.9\,Jy) for features $-$5.57 and $-$4.96\,km\,s$^{-1}$. The relative amplitude of these six features varies from 2 to $>$89. The index of their variability estimated with the $\chi_\mathrm{r}^{2}$ is significant, while the $-$7.68\kms feature is highly variable; its intensity changed by a factor of 6-9 on timescales of 1.5-2.0\,yr. The feature $-$6.10\kms shows relatively weak variability, and its intensity is anti-correlated with all the rest of the features. Figure\,\ref{fig:g78-anti-correlation} compares the light curves of the most and most minor variable features at $-$7.68 and $-$6.10\kms, respectively. There is moderate anti-correlation with the $-$0.55 coefficient for the entire period, but when the interval is limited to around MJD 55600-57100, the coefficient improves to $-$0.63 then declines to $-$0.45. A few episodes of even stronger anti-correlation lasting longer than one month occurred. All but $-$6.10\kms\, features show quasi-periodic ($\sim$3\, weeks) variations in the flux density with the amplitude higher than 20-40\%. There was no time lag between spectral components within an accuracy of 5\,days.

\begin{table}
\caption{Variability characteristics of the spectral features.}
\centering
\begin{tabular}{l c c c c} 
\hline
$V_\mathrm{p}$ & $S_\mathrm{p}$ & $R_\mathrm{a}$ & $\chi_\mathrm{r}^{2}$ & $\chi_{99.9\%}^{2}$\\
(\kms)  & (Jy) &   &  & \\
\hline
$-$8.34 & 3.8   & $>$7.5  &   26.3 & 1.2\\
$-$7.68 & 25.3  & $>$89.2 &   1018.4 \\
$-$7.02 & 16.4  & 11.3 &   57.8\\
$-$6.63 & 24.5  & 13.6 &   78.0 \\
$-$6.10 & 31.0  &  1.9 &   3.5 \\
$-$5.57 & 5.4   & $>$16.4 &   59.3 \\
$-$4.96 & 3.5   & $>$11.4 &   42.7 \\
\hline
\end{tabular}
\tablefoot{$V_\mathrm{p}$ is the peak velocity, $S_\mathrm{p}$ is the average flux density, $R_\mathrm{a}$ is the relative amplitude, $\chi_\mathrm{r}^{2}$ is the reduced $\chi^2$, and $\chi_{99.9\%}^{2}$ is the corresponding value for 99.9\% significance level of variability. 
}
\label{tab:var-prop}
\end{table}

\section{Discussion}
According to most hypotheses, periodic variations of the maser intensity occur in binary systems with a massive star. In such systems, there is a well-defined periodicity with which the interaction of stellar winds near a periastron raises the level of radio background radiation, which may increase the number of seed photons (\citealt{van_der_walt2011,van_der_walt2016}; \citealt{vanderheever2019}).
In close binaries, the accretion rate can periodically vary (\citealt{artymowicz1996}; \citealt{munoz2016}), causing changes in stellar luminosity which modulate the dust temperature and affect the pumping rate (\citealt{araya2010}; \citealt{parfenov2014}). 

A sinusoidal light curve prompts us to consider explaining the cause of variability due to the pulsation of a massive protostar with a high accretion rate (\citealt{inayoshi2013}). For a 2520\,d period, their model implies a luminosity of the central star in IRAS\,20216+4104 of  $\sim$$10^6L_{\odot}$, that is two orders of magnitude higher than what was measured from the spectral energy distribution fit (e.g.,\,\citealt{cesaroni2023}). Therefore, we can rule out this hypothesis.

Spectroscopic and imaging optical observations have shown that 70\% to 90\% massive stars have at least one companion (\citealt{sana2014}; \citealt{lanthermann2023}). A survey of radial velocities of 16 embedded high-mass stars revealed that at least 20\% of the targets are close binaries (\citealt{apai2007}). IRAS\,20216+4104 is probably no exception, as suggested by some hints. X-ray observations revealed that the source is associated with an embedded, possibly very young stellar population (\citealt{montes2015}). \cite{shepherd2000} suggested that the jet precession can be due to tidal interactions between the disk and a companion in a non-coplanar orbit. Interferometric maps of molecular lines implied the Keplerian pattern of the disk. Still, the high energy (513\,K) CH$_3$CN emission presents two distinct peaks, which one located slightly offset from the centre may be due to a lower mass companion orbiting a high-mass central star  (\citealt{cesaroni2005}). The authors have speculated that the observed axial asymmetry of the disc (\citealt{cesaroni2014}) may be caused by a nearby, unseen, low-mass (2\,M$_{\odot}$) companion.

The recent finding of anti-correlated variability of the NIR continuum emission of the N-E and S-E outflow cavities gives another hint at a nearby low-mass companion which may cause the wobbling of the inner disc that alternatively eclipses the two hemispheres of the central massive star (\citealt{massi2023}). Six-epoch NIR photometry suggests a possible periodicity of 12$-$18\,yr (\citealt{massi2023}, their Fig. 4); we note, however, that the NIR continuum emission of a large part (S1 and S2 polygons in their Fig. 2) of the S-E outflow cavity closely follows $\sim$7\,yr periodic behaviour of the 6.7\,GHz methanol maser line (Fig.\,\ref{fig:g78-lc}). In contrast, the emission from the neighbouring region (polygon PD) nearest to the central star shows no similar variation. The brightness variations of the N-W lobe are anti-correlated with the maser intensity with a possible increase of the degree of anti-correlation for the farthest area (polygon N3). As the data are sparse, we can only hypothesize that the NIR continuum emission from the farthest ($\geq$10000\,au) portions of lobes is more strongly (anti-) correlated with the maser emission.
 
\begin{figure}
\centering
\includegraphics[width=1\columnwidth]{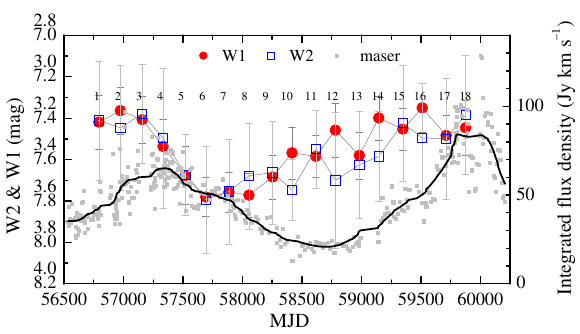}
\caption{NEOWISE lightcurves at 3.4\,$\mu$m (W1) and 4.6\,$\mu$m (W2) of the target superimposed with the 6.7\,GHz maser velocity-integrated flux density. The colours of circles and squares correspond to the average date of NEOWISE exposures in 18 numbered epochs. The bars show the standard deviation. The thick black line shows the average maser light curves. 
\label{fig:g78-mas-ir-lc}}
\end{figure}

\begin{figure}
\centering
\includegraphics[width=1\columnwidth]{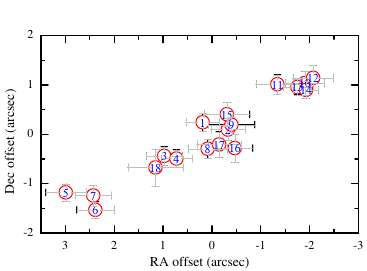}
\caption{Offsets from the overall average position for 18 epochs of NEOWISE observations of IRAS\,20216+4104. The numbered symbols correspond to the epochs in Fig.\,\ref{fig:g78-mas-ir-lc}. The bars indicate the standard errors of the mean. 
\label{fig:g78-ir-offset}}
\end{figure}

\begin{figure}
\centering
\includegraphics[width=1\columnwidth]{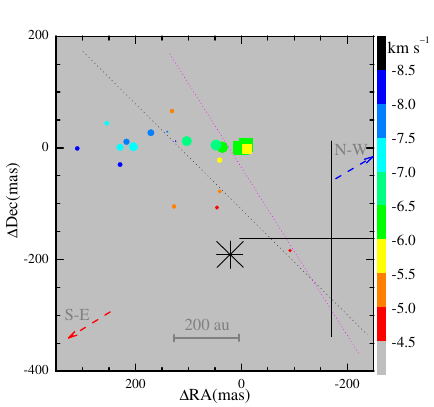}
\caption{Map of the maser spots in IRAS\,20216+4104 obtained on 2012 March 4 with VLA-C (\citealt{hu2016}). The colour and size of symbols correspond to the velocity scale, as shown in the wedge and the logarithm of intensity, respectively. The circles and squares denote the maser spots whose emission synchronously varies with the NIR continuum of the SE and NW lobes (\citealt{massi2023}), respectively. The dashed arrows show the direction of the large-scale ($\sim$10\arcsec) outflow (\citealt{cesaroni2013}). The asterisk marks the nominal position of the protostar derived from the model fit to the water masers (\citealt{moscadelli2011}). The cross marks the position of the 230\,GHz continuum emission peak and its size corresponds to the uncertainty (\citealt{cesaroni2014}). The dotted black line marks the direction of the disc plane (\citealt{cesaroni2005}). The dotted magenta line is orthogonal to the axis of large-scale jet or outflow and divides the maser spot distribution into two parts, which show anti-correlated variability. 
\label{fig:g78-mas-map}}
\end{figure}

To further investigate the source variability, we  used the W1 (3.4\,$\mu$m) and W2 (4.6\,$\mu$m) observations from the NEOWISE (\citealt{mainzer2014}) single exposure database\footnote{https://irsa.ipac.caltech.edu}. The high-quality ($ph\_ qual$ = AA) exposures are only used to obtain the average fluxes at eighteen epochs since May 2014. The W1 and W2 average magnitudes of each epoch superimposed with the maser integrated flux density are shown in Fig.\,\ref{fig:g78-mas-ir-lc}. Significant variability exists with the relative amplitude of 0.28 and 0.37 at 3.4\,$\mu$m and 4.6\,$\mu$m wavelengths, respectively. There is a time lag of 0.12/0.20$P$ between the light curve maxima and minima in the maser and NIR bands, where $P$ denotes the period. High angular-resolution observations have revealed that the maser regions are 390-530\,au away from the central star (\citealt{moscadelli2011}; \citealt{cesaroni2013}; \citealt{aberfelds2023}, see also Fig.\,\ref{fig:g78-mas-map}). This could imply a velocity of 1300-3000\kms for the propagation of infrared radiation, which drives the variations of the maser emission and is one order of magnitude lower than that postulated for thermal heating induced by accretion events (\citealt{burns2020}). Therefore, it is less probable that luminosity bursts of the central star drive the maser variability in IRAS\,20216+4104. The maser light curve of our target appears to be similar to that reported in G331.13$-$0.24, but with a 509\,d period (\citealt{goedhart2004,goedhart2014}) and is not consistent with a simple CWB model (\citealt{van_der_walt2011}). 

Intriguingly, the NEOWISE NIR light centroids show a time-dependent shift of about 5\farcs5 (Fig.\,\ref{fig:g78-ir-offset}) that corresponds to 9000\,au, namely, it is equal to the separation of the brightest inner spots of outflow lobes seen in K$_\mathrm{s}$ band (\citealt{cesaroni2013}). Furthermore, the position angle of this displacement of $-$62\degr\, is in perfect agreement with the outflow axis (\citealt{cesaroni2013}). Epochs 1-3 of NEOWISE observations were likely at maximum NIR radiation and in the last part of the rising phase of the maser, the centroids are located in the middle of the plot. The NIR light centroids started to shift in the S-E direction after the maser maximum, reaching the extreme displacement at epochs 5-7, and then going back to the centre during the decay phase of the maser light curve. In the plateau minimum of the integrated maser emission (epochs 10-14), the centroids were shifted toward the northwest, and then when the maser emission increased in epoch 15, the centroids returned to the central position and started to move toward the southeast. In general, the observed shifts of NIR emission centroids can be an effect of light echo, triggered by luminosity bursts, in outflow cavities oriented outside the plane of the sky (\citealt{caratti2017}; \citealt{stecklum2018}). In the case of IRAS\,20216+4104, the jet or outflow average inclination angle to the plane of the sky is as small as 8$\pm$1\degr\, (\citealt{massi2023}), so the light echo effect is unlikely. We suggest that periodic luminosity variations of the powering protostar might not drive the methanol maser variability in the target.

The waveform light curves were observed in G338.92$-$0.06 (\citealt{goedhart2004,goedhart2014}) and G358.460$-$0.391 (\citealt{maswanganye2015}) and interpreted as an eclipsing effect because they very much resemble that of some eclipsing binary systems (\citealt{maswanganye2015}). They have a symmetric shape around the minima, which lasts 0.04-0.09$P$. Data from the Torun telescope indicate that the curve of IRAS\,20216+4104 has a flat minimum lasting about 0.22$P$. During the minima, the red-shifted emission at a velocity higher than $-$5.6\kms\, drops below 3$\sigma$ detection level, while the emission at lower velocity is commonly above the sensitivity threshold (Fig\,\ref{fig:dyn-spec}). In general, there is a tendency for the blue-shifted ($\leq-$5.6\kms) emission to come from the NE and N sides of the distribution (Fig.\,\ref{fig:g78-mas-map}). The flat and long minima of the emission from these regions are not affected by our sensitivity, in contrast to the red-shifted ($\geq-$5.6\kms) emission from the southern region, which was detected only with the VLA-C (\citealt{hu2016}), but generally unseen in VLBI maps (\citealt{aberfelds2023}). 

A plausible explanation can be that an eclipse of the central star by an inner wobbling disk can modulate the scattered light in the outflow lobes. Moreover, this eclipsing can vary the maser transition's pumping rate by cyclic shading portions of the dust disk and indirectly affect the maser intensity. The inner disk cannot eclipse the maser itself since it is 390$-$530\,au from the massive star in the N-E part of the gas disk (Fig.\,\ref{fig:g78-mas-map}). The shadowed portions rotate over the dust disk surface and prograde with wobbling disk motion. A simple visualisation shows that a maximum of NIR radiation seen by the observer occurs for PA$\simeq$57\degr of the major axis of the wobbling disk (Fig.\,\ref{fig:scheme}b).  Since the disk rotates counterclockwise (\citealt{cesaroni2014}), this maximum of NIR radiation seen by NEOWISE precedes the maximum pumping radiation seen through the maser region (Fig.\,\ref{fig:scheme}, c). Following the discussion in \cite{massi2023}, we refine the system's parameters. Assuming the masses of primary and secondary protostars are $12M_{\odot}$ and $2M_{\odot}$, respectively (\citealt{chen2016}; \citealt{cesaroni2014}) and with the secondary's orbit inclination of 30\degr\, relative to the disk plane and orbital radius of 14\,au, we obtain a disk radius of 12\,au. Such values fall well within the ranges observed for pre-main sequence stars (\citealt{terquem1999}). Interpretations suggesting the periodic variability of the maser emission is attributed to an eclipsing effect are supported by a characteristic feature of the maser light curve in IRAS\,20216+4104, namely:\ the rising and falling edge adjacent to the minimum is relatively symmetric standard for eclipsing binary systems.

The maser emission observed with the VLA-C in March 2012 (\citealt{hu2016}) is elongated in the NE$-$SW direction with a velocity gradient (Fig.\,\ref{fig:g78-mas-map}) and it partially overlaps with the NE part of the gas disk of $\sim$1\farcs5 in size, as traced by the thermal lines of CH$_3$CN at 230\,GHz (\citealt{cesaroni2014}). The extremely northwesterly components around $-$6.1\kms are marked as the squares in Fig.\,\ref{fig:g78-mas-map}, lying about 90\,au away from the rest of the maser clouds; they do not follow a general trend in terms of their variations. This is probably because they could be located in the outer regions of the jet as indicated a  proper motion study (\citealt{moscadelli2011}). Consequently, these regions may not have been exposed (or only weakly) to eclipse effects by the wobbling of the inner disk. 

\section{Conclusions}
In this paper, we characterize the long-term variability of the 6.7\,GHz methanol maser emission of the high-mass protostar IRAS\,20126+4104 using both new and archival data from a 31\,yr period. The flux density of all but one spectral feature shows a variation with a period of 2520\,d ($\sim$7\,yr). This is the longest observed period, exceeding twice the period of the previous known longest periodic source. The maser variations strongly correlate with the NIR continuum emission of the south-eastern outflow cavity measured six times over 20\,yr by \cite{massi2023}. Data from 18 NEOWISE observations over 8.5\,yr indicate variations of the IR emission, whose peak precedes by one-fifth of the period the maser maximum. The eclipsing effect of the central star by a wobbling inner disk caused by a nearby orbiting low-mass companion can possibly explain the characteristics of both maser and infrared variability. 

This serendipitous finding of long-term periodic variability of the methanol maser in the extensively studied massive star with the disk or jet system could trigger multi-band monitoring over a flare cycle to constrain maser models as well as contribute to our understanding of binary and multiple systems with a massive primary.

\section{Acknowledgements}
The 32\,m radio telescope is operated by the Institute of Astronomy, Nicolaus Copernicus University, and supported by the Polish Ministry of Science and Higher Education SpUB grant. We thank the staff and students for their assistance with the observations. M.S. and his Torun collaborators acknowledge support from the National Science Centre, Poland, through grant no. 2021/43/B/ST9/02008.
The research has made use of the SIMBAD database, operated at CDS (Strasbourg, France), as well as NASA's Astrophysics Data System Bibliographic Services.
This publication also makes use of data products from NEOWISE, which is a project of the Jet Propulsion Laboratory/California Institute of Technology, funded by the Planetary Science Division of the National Aeronautics and Space Administration. 

%

\bibliography{librarian}{}

\begin{thebibliography}{55}
\expandafter\ifx\csname natexlab\endcsname\relax\def\natexlab#1{#1}\fi

\bibitem[{{Aberfelds} {et~al.}(2023){Aberfelds}, {Bartkiewicz}, {Szymczak},
  {{\v{S}}teinbergs}, {Surcis}, {Kobak}, {Durjasz}, \&
  {Shmeld}}]{aberfelds2023}
{Aberfelds}, A., {Bartkiewicz}, A., {Szymczak}, M., {et~al.} 2023, \mnras, 524,
  599

\bibitem[{{Apai} {et~al.}(2007){Apai}, {Bik}, {Kaper}, {Henning}, \&
  {Zinnecker}}]{apai2007}
{Apai}, D., {Bik}, A., {Kaper}, L., {Henning}, T., \& {Zinnecker}, H. 2007,
  \apj, 655, 484

\bibitem[{{Araya} {et~al.}(2010){Araya}, {Hofner}, {Goss}, {Kurtz}, {Richards},
  {Linz}, {Olmi}, \& {Sewi{\l}o}}]{araya2010}
{Araya}, E.~D., {Hofner}, P., {Goss}, W.~M., {et~al.} 2010, \apjl, 717, L133

\bibitem[{{Artymowicz} \& {Lubow}(1996)}]{artymowicz1996}
{Artymowicz}, P. \& {Lubow}, S.~H. 1996, \apjl, 467, L77

\bibitem[{{Burns} {et~al.}(2020){Burns}, {Sugiyama}, {Hirota}, {Kim},
  {Sobolev}, {Stecklum}, {MacLeod}, {Yonekura}, {Olech}, {Orosz}, {Ellingsen},
  {Hyland}, {Caratti o Garatti}, {Brogan}, {Hunter}, {Phillips}, {van den
  Heever}, {Eisl{\"o}ffel}, {Linz}, {Surcis}, {Chibueze}, {Baan}, \&
  {Kramer}}]{burns2020}
{Burns}, R.~A., {Sugiyama}, K., {Hirota}, T., {et~al.} 2020, Nature Astronomy,
  4, 506

\bibitem[{{Caratti o Garatti} {et~al.}(2017){Caratti o Garatti}, {Stecklum},
  {Garcia Lopez}, {Eisl{\"o}ffel}, {Ray}, {Sanna}, {Cesaroni}, {Walmsley},
  {Oudmaijer}, {de Wit}, {Moscadelli}, {Greiner}, {Krabbe}, {Fischer}, {Klein},
  \& {Iba{\~n}ez}}]{caratti2017}
{Caratti o Garatti}, A., {Stecklum}, B., {Garcia Lopez}, R., {et~al.} 2017,
  Nature Physics, 13, 276

\bibitem[{{Cesaroni} {et~al.}(2023){Cesaroni}, {Faustini}, {Galli},
  {Lorenzani}, {Molinari}, \& {Testi}}]{cesaroni2023}
{Cesaroni}, R., {Faustini}, F., {Galli}, D., {et~al.} 2023, \aap, 671, A126

\bibitem[{{Cesaroni} {et~al.}(2007){Cesaroni}, {Galli}, {Lodato}, {Walmsley},
  \& {Zhang}}]{cesaroni2007}
{Cesaroni}, R., {Galli}, D., {Lodato}, G., {Walmsley}, C.~M., \& {Zhang}, Q.
  2007, in Protostars and Planets V, ed. B.~{Reipurth}, D.~{Jewitt}, \&
  K.~{Keil}, 197

\bibitem[{{Cesaroni} {et~al.}(2014){Cesaroni}, {Galli}, {Neri}, \&
  {Walmsley}}]{cesaroni2014}
{Cesaroni}, R., {Galli}, D., {Neri}, R., \& {Walmsley}, C.~M. 2014, \aap, 566,
  A73

\bibitem[{{Cesaroni} {et~al.}(2013){Cesaroni}, {Massi}, {Arcidiacono},
  {Beltr{\'a}n}, {McCarthy}, {Kulesa}, {Boutsia}, {Paris},
  {Quir{\'o}s-Pacheco}, \& {Xompero}}]{cesaroni2013}
{Cesaroni}, R., {Massi}, F., {Arcidiacono}, C., {et~al.} 2013, \aap, 549, A146

\bibitem[{{Cesaroni} {et~al.}(2005){Cesaroni}, {Neri}, {Olmi}, {Testi},
  {Walmsley}, \& {Hofner}}]{cesaroni2005}
{Cesaroni}, R., {Neri}, R., {Olmi}, L., {et~al.} 2005, \aap, 434, 1039

\bibitem[{{Chen} {et~al.}(2016){Chen}, {Keto}, {Zhang}, {Sridharan}, {Liu}, \&
  {Su}}]{chen2016}
{Chen}, H.-R.~V., {Keto}, E., {Zhang}, Q., {et~al.} 2016, \apj, 823, 125

\bibitem[{{Felli} {et~al.}(2007){Felli}, {Brand}, {Cesaroni}, {Codella},
  {Comoretto}, {Di Franco}, {Massi}, {Moscadelli}, {Nesti}, {Olmi}, {Palagi},
  {Panella}, \& {Valdettaro}}]{felli2007}
{Felli}, M., {Brand}, J., {Cesaroni}, R., {et~al.} 2007, \aap, 476, 373

\bibitem[{{Fontani} {et~al.}(2010){Fontani}, {Cesaroni}, \&
  {Furuya}}]{fontani2010}
{Fontani}, F., {Cesaroni}, R., \& {Furuya}, R.~S. 2010, \aap, 517, A56

\bibitem[{{Fujisawa} {et~al.}(2014){Fujisawa}, {Takase}, {Kimura}, {Aoki},
  {Nagadomi}, {Shimomura}, {Sugiyama}, {Motogi}, {Niinuma}, {Hirota}, \&
  {Yonekura}}]{fujisawa2014}
{Fujisawa}, K., {Takase}, G., {Kimura}, S., {et~al.} 2014, \pasj, 66, 78

\bibitem[{{Goedhart} {et~al.}(2003){Goedhart}, {Gaylard}, \& {van der
  Walt}}]{goedhart2003}
{Goedhart}, S., {Gaylard}, M.~J., \& {van der Walt}, D.~J. 2003, \mnras, 339,
  L33

\bibitem[{{Goedhart} {et~al.}(2004){Goedhart}, {Gaylard}, \& {van der
  Walt}}]{goedhart2004}
{Goedhart}, S., {Gaylard}, M.~J., \& {van der Walt}, D.~J. 2004, \mnras, 355,
  553

\bibitem[{{Goedhart} {et~al.}(2009){Goedhart}, {Langa}, {Gaylard}, \& {van der
  Walt}}]{goedhart2009}
{Goedhart}, S., {Langa}, M.~C., {Gaylard}, M.~J., \& {van der Walt}, D.~J.
  2009, \mnras, 398, 995

\bibitem[{{Goedhart} {et~al.}(2014){Goedhart}, {Maswanganye}, {Gaylard}, \&
  {van der Walt}}]{goedhart2014}
{Goedhart}, S., {Maswanganye}, J.~P., {Gaylard}, M.~J., \& {van der Walt},
  D.~J. 2014, \mnras, 437, 1808

\bibitem[{{Gray} {et~al.}(2020){Gray}, {Etoka}, {Travis}, \&
  {Pimpanuwat}}]{gray2020}
{Gray}, M.~D., {Etoka}, S., {Travis}, A., \& {Pimpanuwat}, B. 2020, \mnras,
  493, 2472

\bibitem[{{Hu} {et~al.}(2016){Hu}, {Menten}, {Wu}, {Bartkiewicz}, {Rygl},
  {Reid}, {Urquhart}, \& {Zheng}}]{hu2016}
{Hu}, B., {Menten}, K.~M., {Wu}, Y., {et~al.} 2016, \apj, 833, 18

\bibitem[{{Inayoshi} {et~al.}(2013){Inayoshi}, {Hosokawa}, \&
  {Omukai}}]{inayoshi2013}
{Inayoshi}, K., {Hosokawa}, T., \& {Omukai}, K. 2013, \mnras, 431, 3036

\bibitem[{{Johnston} {et~al.}(2011){Johnston}, {Keto}, {Robitaille}, \&
  {Wood}}]{johnston2011}
{Johnston}, K.~G., {Keto}, E., {Robitaille}, T.~P., \& {Wood}, K. 2011, \mnras,
  415, 2953

\bibitem[{{Lanthermann} {et~al.}(2023){Lanthermann}, {Le Bouquin}, {Sana},
  {M{\'e}rand}, {Monnier}, {Perraut}, {Frost}, {Mahy}, {Gosset}, {De Becker},
  {Kraus}, {Anugu}, {Davies}, {Ennis}, {Gardner}, {Labdon}, {Setterholm}, {ten
  Brummelaar}, \& {Schaefer}}]{lanthermann2023}
{Lanthermann}, C., {Le Bouquin}, J.~B., {Sana}, H., {et~al.} 2023, \aap, 672,
  A6

\bibitem[{{MacLeod} \& {Gaylard}(1992)}]{macleod1992}
{MacLeod}, G.~C. \& {Gaylard}, M.~J. 1992, \mnras, 256, 519

\bibitem[{{Mainzer} {et~al.}(2014){Mainzer}, {Bauer}, {Cutri}, {Grav},
  {Masiero}, {Beck}, {Clarkson}, {Conrow}, {Dailey}, {Eisenhardt}, {Fabinsky},
  {Fajardo-Acosta}, {Fowler}, {Gelino}, {Grillmair}, {Heinrichsen}, {Kendall},
  {Kirkpatrick}, {Liu}, {Masci}, {McCallon}, {Nugent}, {Papin}, {Rice},
  {Royer}, {Ryan}, {Sevilla}, {Sonnett}, {Stevenson}, {Thompson}, {Wheelock},
  {Wiemer}, {Wittman}, {Wright}, \& {Yan}}]{mainzer2014}
{Mainzer}, A., {Bauer}, J., {Cutri}, R.~M., {et~al.} 2014, \apj, 792, 30

\bibitem[{{Massi} {et~al.}(2023){Massi}, {Caratti o Garatti}, {Cesaroni},
  {Sridharan}, {Ghose}, {Pinna}, {Beltr{\'a}n}, {Leurini}, {Moscadelli},
  {Sanna}, {Agapito}, {Briguglio}, {Christou}, {Esposito}, {Mazzoni}, {Miller},
  {Plantet}, {Power}, {Puglisi}, {Rossi}, {Rothberg}, {Taylor}, \&
  {Veillet}}]{massi2023}
{Massi}, F., {Caratti o Garatti}, A., {Cesaroni}, R., {et~al.} 2023, \aap, 672,
  A113

\bibitem[{{Maswanganye} {et~al.}(2015){Maswanganye}, {Gaylard}, {Goedhart},
  {Walt}, \& {Booth}}]{maswanganye2015}
{Maswanganye}, J.~P., {Gaylard}, M.~J., {Goedhart}, S., {Walt}, D.~J. v.~d., \&
  {Booth}, R.~S. 2015, \mnras, 446, 2730

\bibitem[{{Maswanganye} {et~al.}(2016){Maswanganye}, {van der Walt},
  {Goedhart}, \& {Gaylard}}]{maswanganye2016}
{Maswanganye}, J.~P., {van der Walt}, D.~J., {Goedhart}, S., \& {Gaylard},
  M.~J. 2016, \mnras, 456, 4335

\bibitem[{{Montes} {et~al.}(2015){Montes}, {Hofner}, {Anderson}, \&
  {Rosero}}]{montes2015}
{Montes}, V.~A., {Hofner}, P., {Anderson}, C., \& {Rosero}, V. 2015, \apjs,
  219, 41

\bibitem[{{Moscadelli} {et~al.}(2011){Moscadelli}, {Cesaroni}, {Rioja},
  {Dodson}, \& {Reid}}]{moscadelli2011}
{Moscadelli}, L., {Cesaroni}, R., {Rioja}, M.~J., {Dodson}, R., \& {Reid},
  M.~J. 2011, \aap, 526, A66

\bibitem[{{Mu{\~n}oz} \& {Lai}(2016)}]{munoz2016}
{Mu{\~n}oz}, D.~J. \& {Lai}, D. 2016, \apj, 827, 43

\bibitem[{{Olech} {et~al.}(2022){Olech}, {Durjasz}, {Szymczak}, \&
  {Bartkiewicz}}]{olech2022}
{Olech}, M., {Durjasz}, M., {Szymczak}, M., \& {Bartkiewicz}, A. 2022, \aap,
  661, A114

\bibitem[{{Olech} {et~al.}(2019){Olech}, {Szymczak}, {Wolak}, {Sarniak}, \&
  {Bartkiewicz}}]{olech2019}
{Olech}, M., {Szymczak}, M., {Wolak}, P., {Sarniak}, R., \& {Bartkiewicz}, A.
  2019, \mnras, 486, 1236

\bibitem[{{Parfenov} \& {Sobolev}(2014)}]{parfenov2014}
{Parfenov}, S.~Y. \& {Sobolev}, A.~M. 2014, \mnras, 444, 620

\bibitem[{{Proven-Adzri} {et~al.}(2019){Proven-Adzri}, {MacLeod}, {Heever},
  {Hoare}, {Kuditcher}, \& {Goedhart}}]{proven-adzri2019}
{Proven-Adzri}, E., {MacLeod}, G.~C., {Heever}, S.~P. v.~d., {et~al.} 2019,
  \mnras, 487, 2407

\bibitem[{{Sana} {et~al.}(2014){Sana}, {Le Bouquin}, {Lacour}, {Berger},
  {Duvert}, {Gauchet}, {Norris}, {Olofsson}, {Pickel}, {Zins}, {Absil}, {de
  Koter}, {Kratter}, {Schnurr}, \& {Zinnecker}}]{sana2014}
{Sana}, H., {Le Bouquin}, J.~B., {Lacour}, S., {et~al.} 2014, \apjs, 215, 15

\bibitem[{{Shepherd} {et~al.}(2000){Shepherd}, {Yu}, {Bally}, \&
  {Testi}}]{shepherd2000}
{Shepherd}, D.~S., {Yu}, K.~C., {Bally}, J., \& {Testi}, L. 2000, \apj, 535,
  833

\bibitem[{{Slysh} {et~al.}(1999){Slysh}, {Val'tts}, {Kalenskii}, {Voronkov},
  {Palagi}, {Tofani}, \& {Catarzi}}]{slysh1999}
{Slysh}, V.~I., {Val'tts}, I.~E., {Kalenskii}, S.~V., {et~al.} 1999, \aaps,
  134, 115

\bibitem[{{Stecklum} {et~al.}(2018){Stecklum}, {Caratti o Garatti}, {Hodapp},
  {Linz}, {Moscadelli}, \& {Sanna}}]{stecklum2018}
{Stecklum}, B., {Caratti o Garatti}, A., {Hodapp}, K., {et~al.} 2018, in
  Astrophysical Masers: Unlocking the Mysteries of the Universe, ed.
  A.~{Tarchi}, M.~J. {Reid}, \& P.~{Castangia}, Vol. 336, 37--40

\bibitem[{{Sugiyama} {et~al.}(2017){Sugiyama}, {Nagase}, {Yonekura}, {Momose},
  {Yasui}, {Saito}, {Motogi}, {Honma}, {Hachisuka}, {Matsumoto}, {Uchiyama}, \&
  {Fujisawa}}]{sugiyama2017}
{Sugiyama}, K., {Nagase}, K., {Yonekura}, Y., {et~al.} 2017, \pasj, 69, 59

\bibitem[{{Sugiyama} {et~al.}(2015){Sugiyama}, {Yonekura}, {Motogi}, {Saito},
  {Fujisawa}, {Ishii}, {Momose}, {Honma}, {Tazaki}, {Tanaka}, {Hosokawa},
  {Uchiyama}, \& {Inayoshi}}]{sugiyama2015}
{Sugiyama}, K., {Yonekura}, Y., {Motogi}, K., {et~al.} 2015, Publication of
  Korean Astronomical Society, 30, 129

\bibitem[{{Szymczak} {et~al.}(2000){Szymczak}, {Hrynek}, \&
  {Kus}}]{szymczak2000}
{Szymczak}, M., {Hrynek}, G., \& {Kus}, A.~J. 2000, \aaps, 143, 269

\bibitem[{{Szymczak} {et~al.}(2018){Szymczak}, {Olech}, {Sarniak}, {Wolak}, \&
  {Bartkiewicz}}]{szymczak2018a}
{Szymczak}, M., {Olech}, M., {Sarniak}, R., {Wolak}, P., \& {Bartkiewicz}, A.
  2018, \mnras, 474, 219

\bibitem[{{Szymczak} {et~al.}(2016){Szymczak}, {Olech}, {Wolak}, {Bartkiewicz},
  \& {Gawro{\'n}ski}}]{szymczak_2016}
{Szymczak}, M., {Olech}, M., {Wolak}, P., {Bartkiewicz}, A., \&
  {Gawro{\'n}ski}, M. 2016, \mnras, 459, L56

\bibitem[{{Szymczak} {et~al.}(2015){Szymczak}, {Wolak}, \&
  {Bartkiewicz}}]{szymczak2015}
{Szymczak}, M., {Wolak}, P., \& {Bartkiewicz}, A. 2015, \mnras, 448, 2284

\bibitem[{{Szymczak} {et~al.}(2012){Szymczak}, {Wolak}, {Bartkiewicz}, \&
  {Borkowski}}]{szymczak2012}
{Szymczak}, M., {Wolak}, P., {Bartkiewicz}, A., \& {Borkowski}, K.~M. 2012,
  Astronomische Nachrichten, 333, 634

\bibitem[{{Szymczak} {et~al.}(2011){Szymczak}, {Wolak}, {Bartkiewicz}, \& {van
  Langevelde}}]{szymczak2011}
{Szymczak}, M., {Wolak}, P., {Bartkiewicz}, A., \& {van Langevelde}, H.~J.
  2011, \aap, 531, L3

\bibitem[{{Tanabe} {et~al.}(2023){Tanabe}, {Yonekura}, \&
  {MacLeod}}]{tanabe2023}
{Tanabe}, Y., {Yonekura}, Y., \& {MacLeod}, G.~C. 2023, \pasj, 75, 351

\bibitem[{{Terquem} {et~al.}(1999){Terquem}, {Eisl{\"o}ffel}, {Papaloizou}, \&
  {Nelson}}]{terquem1999}
{Terquem}, C., {Eisl{\"o}ffel}, J., {Papaloizou}, J.~C.~B., \& {Nelson}, R.~P.
  1999, \apjl, 512, L131

\bibitem[{{van den Heever} {et~al.}(2019){van den Heever}, {van der Walt},
  {Pittard}, \& {Hoare}}]{vanderheever2019}
{van den Heever}, S.~P., {van der Walt}, D.~J., {Pittard}, J.~M., \& {Hoare},
  M.~G. 2019, \mnras, 485, 2759

\bibitem[{{van der Walt}(2011)}]{van_der_walt2011}
{van der Walt}, D.~J. 2011, \aj, 141, 152

\bibitem[{{van der Walt} {et~al.}(2016){van der Walt}, {Maswanganye}, {Etoka},
  {Goedhart}, \& {van den Heever}}]{van_der_walt2016}
{van der Walt}, D.~J., {Maswanganye}, J.~P., {Etoka}, S., {Goedhart}, S., \&
  {van den Heever}, S.~P. 2016, \aap, 588, A47

\bibitem[{{Vlemmings}(2008)}]{vlemmings2008}
{Vlemmings}, W.~H.~T. 2008, \aap, 484, 773

\bibitem[{{Yang} {et~al.}(2017){Yang}, {Chen}, {Shen}, {Li}, {Wang}, {Jiang},
  {Li}, {Dong}, {Wu}, {Qiao}, \& {Ren}}]{yang_2017}
{Yang}, K., {Chen}, X., {Shen}, Z.-Q., {et~al.} 2017, \apj, 846, 160

\end{thebibliography}
\bibliographystyle{aa}

\begin{appendix}

\section{Additional table and figures}
\begin{figure*}
\centering
\includegraphics[width=2\columnwidth]{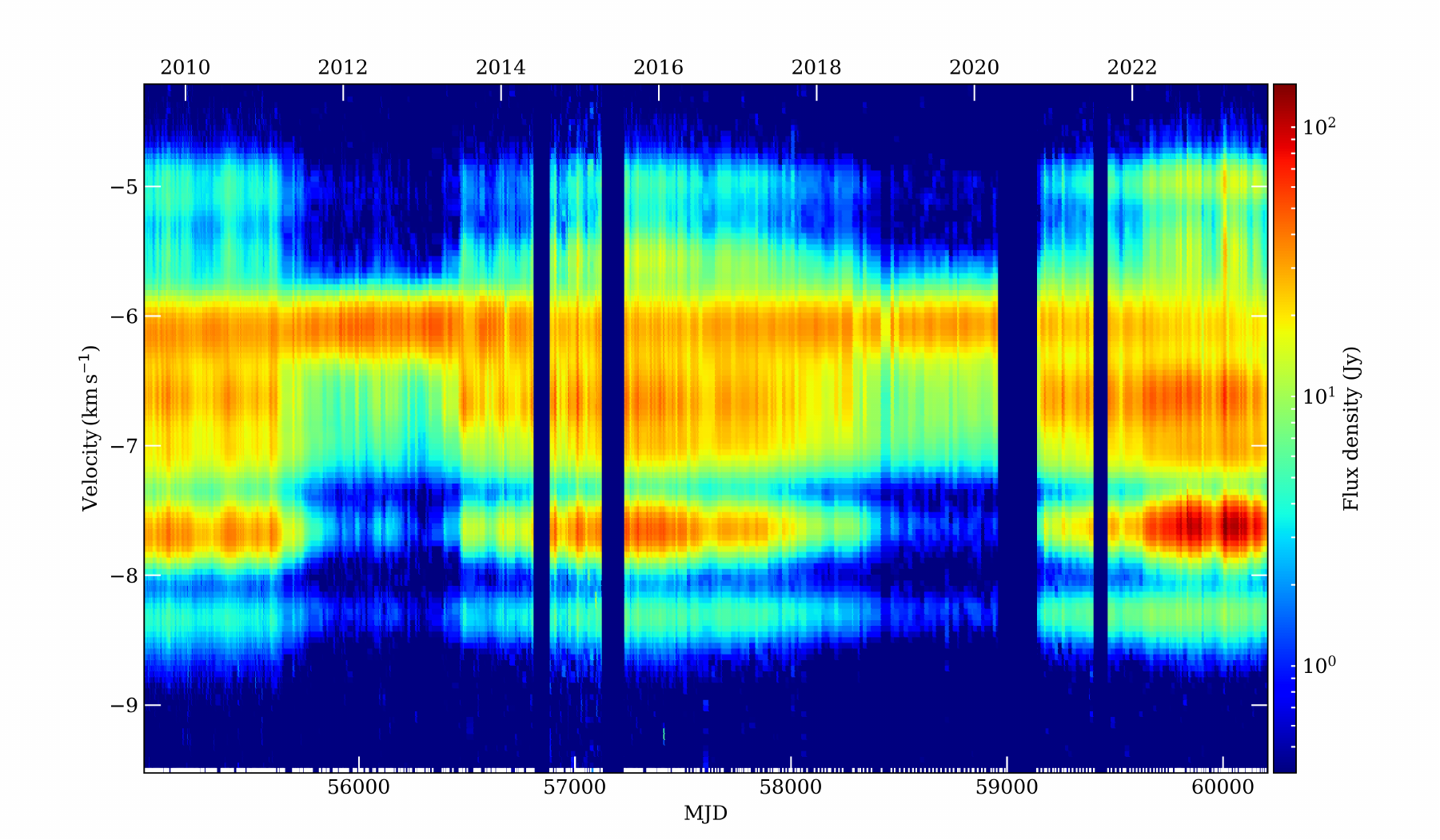}
\caption{Dynamic spectrum of the 6.7 GHz methanol maser emission of IRAS\,20216+4104. Velocity is measured relative to the local standard of rest. The vertical bars in the bottom ordinate correspond to the dates of the observed spectra. \label{fig:dyn-spec}}
\end{figure*}

\begin{figure*}
\centering
\includegraphics[width=1.5\columnwidth]{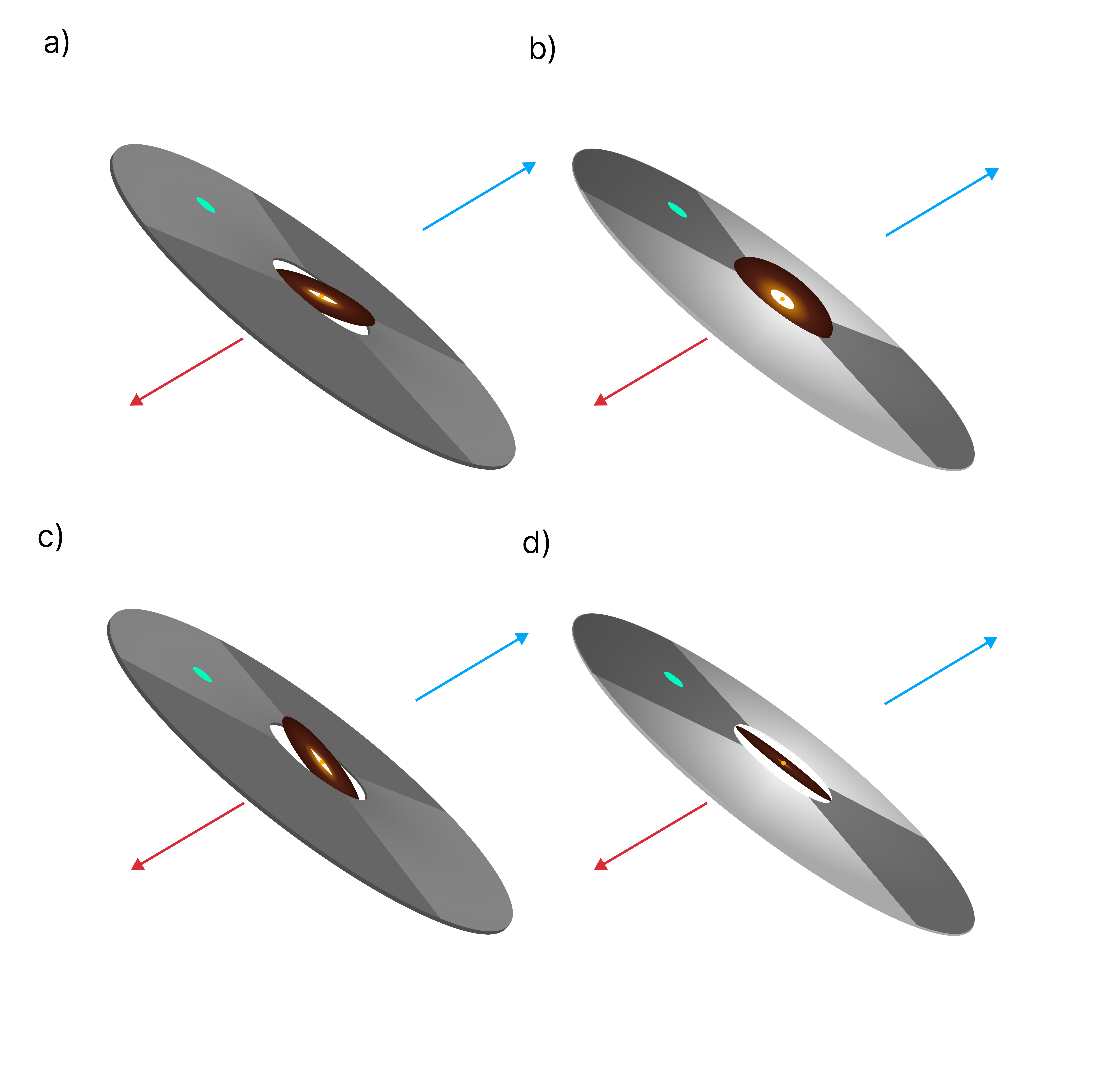}
\caption{Sketch of variability scenario in IRAS\,20216+4104. The inner part of the disk of massive protostar wobbles due to the interaction of a nearby low-mass companion with the orbital period of $\sim$14\,yr and alternatively eclipses the two hemispheres of the central protostar. The eclipse modulates the scattered light in the outflow lobes (blue and red arrows) and the dust temperature, causing the maser (green ellipse) intensity variations. The shades of grey encode the relative irradiance of the outer disk. The orientation of the major axis of the disk and outflow lobes and the maser location is from the observations (see the caption of Fig.\,\ref{fig:g78-mas-map}), but the sizes of the model components are not shown to scale. 
\label{fig:scheme}}
\end{figure*}

\begin{table}
\caption{Integrated flux density of 6.7\,GHz methanol maser, $S_\mathrm{i}$, in IRAS20216+4104 from the literature.}
\begin{tabular}{l c c c c c} 
\hline
Epoch   & $S_\mathrm{i}$   & Telescope & Reference \\
MJD     &    (Jy\kms)      &           &           \\
\hline
48556$\pm$14 & 4   & 26m & 1 \\
49810$\pm$7  & 70  & 32m & 2 \\
51271      & 18.3 & 32m & 3 \\
52774$\pm$15 & 42.9 & 100m & 4 \\
54416      & 63   & 100m & 5 \\
54815      & 51.1 & 32m  & 6 \\
56004      & 31.2 & VLA-C & 7 \\
57548      & 63.1 & 65m   & 8 \\
\hline
\end{tabular}\\
\tablebib{
(1) \citet{macleod1992}; (2) \citet{slysh1999}; (3) \citet{szymczak2000}; (4) \citet{fontani2010};
(5) \citet{vlemmings2008}; (6) \citet{szymczak2012}; (7) \citet{hu2016}; (8) \citet{yang_2017}.
}
\label{tab:data-literature}
\end{table}

\end{appendix}
\end{document}